\documentstyle[floats,aps]{revtex}

\begin{document}
\draft
\catcode`\@=11
\catcode`\@=12
\twocolumn[\hsize\textwidth\columnwidth\hsize\csname%
@twocolumnfalse\endcsname
%%%%%%%%%%%%%%%%%%%%%%%
\title{Mean-Field and Perturbation Theory of
Vortex-like Composite Fermions}
%%%%%%%%%%%%%%%%%%%%%%%
\author{Yong-Shi Wu and Yue Yu}
\address{Department of Physics, University of Utah, 
Salt Lake City, UT 84112}

\date{Received: July 31, 1996}
%\receipt{}
\maketitle
\begin{abstract}
We develop a field theory for a partially filled 
Landau level based on composite fermions with   
a finite vortex core, whose mean-field states are 
exactly those described by well-tested trial wave 
functions. Despite non-orthogonality of free
composite-fermion states and non-Hermiticity of the 
mean-field Hamiltonian, a consistent perturbation 
theory is formulated and the mean-field Fermi 
sea at half filling is shown to be stable. 
While the low-energy and long-distance physics is the 
same as in the Chern-Simons fermion theory, new 
physics is expected to show up for larger wave vectors.

\end{abstract}

]

Physics of interacting planar electrons in a partially 
filled lowest Landau level (LLL) in a strong magnetic
field $B$ is surprisingly rich. For example, at the 
filling factor $\nu=1/\tilde{\phi}$, depending on whether 
$\tilde{\phi}$ is an odd or even integer, the system may be 
either in an incompressible \cite{Lau,Halp,Hald} or 
compressible \cite{HLR} fluid state, exhibiting 
the quantized \cite{FQHE} or unquantized \cite{Half} 
Hall effect respectively. The corresponding ground 
states are known to be described by 
numerically well-tested {\it trial} wave functions, 
such as the Laughlin \cite{Lau} and Rezayi-Read
\cite{RR} ones. Physically, these wave functions
can be understood in terms of the composite 
fermion (CF) scenario \cite{Jain,Read1}: Namely 
at fillings near $\nu=1/\tilde{\phi}$, with 
$\tilde{\phi}$ even, a quasiparticle is an 
electron bound to a $\tilde{\phi}$-fold 
{\it vortex} in the electron fluid \cite{HalWu}, 
called a CF, and the strongly correlated fluid 
itself can be described as a collection of such 
{\it weakly interacting} quasiparticles 
in a {\it weaker} effective magnetic 
field $\Delta B= B-B_{\tilde{\phi}}$ with 
$B_{\tilde{\phi}}\equiv \tilde{\phi} n_{e} 
\phi_0$. (Here $n_{e}$ is the average electron 
density, $\phi_0$ the flux quantum.) For 
example, if the residue interactions are 
ignored, the ground state at $\nu=1/2$ 
corresponds to a Fermi sea of CF's, while
that at $\nu=1/3$ to a completely filled LLL of
CF's in $\Delta B$.

In this Letter, we will report on a local
field theoretical approach that allows a systematic 
improvement of these trial ground states, which 
are managed to appear as the {\it mean-field} 
ground states. Our approach is based on a full 
realization of the CF scenario, in which the 
CF has a {\it finite vortex core}, improving
the usual Chern-Simons fermion (CSF) theory 
\cite{LF,HLR,KalZh} in which an 
infinitesimally thin $\tilde{\phi}$-flux is 
attached to each electron. 
The inclusion of a finite vortex core is 
implemented by a {\it non-unitary} transformation, 
which makes CF states with definite momenta 
{\it non-orthogonal} to each other and
the mean-field and perturbed Hamiltonians
{\it non-Hermitian}. Despite these seemingly 
troublesome features, it is shown possible to 
formulate a {\it consistent} perturbation theory,
in which the mean-field ground state, a 
Fermi sea of CF's, at $\nu=1/\tilde{\phi}$ 
is perturbatively stable. We have also 
computed the density-current 
response functions in the random phase 
approximation (RPA), and explicitly verified that 
for small wave vectors they indeed agree with 
those in the CSF theory\cite{HLR}.
New physics due to the finite vortex core 
is expected to show up for larger wave vectors.

Generalizing a recent composite-boson
construction\cite{Raj},
we introduce the CF field operators by
(with $\tilde{\phi}$ {\it even})
\begin{eqnarray}
\Phi(\vec{x}) =e^{-J_{\tilde\phi}
(\vec{x})}\psi(\vec{x}),
\;\;\;
\Pi(\vec{x}) =\psi^\dagger(\vec{x})
e^{J_{\tilde\phi}(\vec{x})},
\label{PhiPi}
\end{eqnarray}
where $\psi(\vec{x})$ is the (spinless) electron 
field operator, and 
\begin{equation}
J_{\tilde\phi}(\vec{x})= {\tilde\phi}\int d^2x'
\; \rho(\vec{x}')\log(z-z') -|z|^2/4l^2_ {\tilde\phi},
\label{J}
\end{equation}
with $l_{\tilde{\phi}}$ the magnetic length
in $B_{\tilde{\phi}}$ and $z=x+iy$.
The usual CSF transform\cite{HLR} 
contains only the imaginary part 
of $\tilde{\phi} \log(z-z')$, which describes 
the phases due to a vortex with vorticity 
$\tilde{\phi}$ bound to an electron at $z'$. 
We have included the real part too, 
incorporating a finite vortex core and making  
the transformation {\it non-unitary} \cite{MaZh}. 
Using 
\begin{eqnarray}
e^{-J_{\tilde\phi}(\vec{x})}\psi(\vec{x}')
&=&(z-z')^{\tilde\phi}\psi(\vec{x}')
e^{-J_{\tilde\phi}(\vec{x})},\nonumber\\
\psi^\dagger(\vec{x}')e^{-J_{\tilde\phi}(\vec{x})}
&=&(z-z')^{\tilde\phi}e^{-J_{\tilde\phi}(\vec{x})}
\psi^\dagger(\vec{x}'),
\label{2.12}
\end{eqnarray}
it is easy to verify $\{\Phi(\vec{x}),\Phi(\vec{x}')\}
=\{\Pi(\vec{x}),\Pi(\vec{x}')\}=0$, and $\{\Phi(\vec{x}),
\Pi(\vec{x}')\}=\delta^{(2)}(\vec{x}-\vec{x}')$.
Obviously $\Phi$ and $\Pi$ are 
not Hermitian conjugate: $\Pi=\Phi^\dagger e^{J_{\tilde\phi}
+J_{\tilde\phi}\dagger}$. Notice that $\rho(\vec{x})=
\psi^\dagger(\vec{x})\psi(\vec{x})
=\Pi(\vec{x})\Phi(\vec{x})$ and $[\int d^2x' \rho,
\Pi(\vec{x})]=\Pi(\vec{x})$. 
Thus the CF density is the same as the electron
density, and $\Pi$ creates a CF 
while $\Phi$ annihilates one. 

In terms of CF, the usual electron Hamiltonian 
reads
\begin{eqnarray}
H&=&-\frac{1}{2m_b} \int d^2x\, \Pi(\vec{x})
(\nabla+i{\vec A}-i{\vec v}_{\tilde\phi})^2
\, \Phi(\vec{x})\nonumber\\
&&\;\;\;+\,\frac{1}{2}\int d^2x d^2x'\,
\delta\rho(\vec{x})\, V(\vec{x}-\vec{x}')
\, \delta\rho(\vec{x}'),
\label{CFHam}
\end{eqnarray}
where $m_b$ is the electron band mass; 
$\delta\rho =\Pi\Phi - n_e$, and
${\vec v}_{\tilde\phi}(\vec{x})
\equiv i\nabla J_{\tilde\phi}
={\vec a}(\vec{x})+i\hat{n}\times{\vec a}
(\vec{x})-i{\vec x}/2l_{\tilde\phi}^2$; 
$\hat{n}$ is a unit vector 
perpendicular to the plane and $\vec{a}$
is the usual Chern-Simons gauge field,
given by
\begin{equation}
{\vec a}(\vec{x})= \tilde{\phi}\,
\int d^2x'\, \rho(\vec{x}') 
\; (\hat{n}\times \vec{x})/ |\vec{x}|^2 
\label{CS}
\end{equation}
in the gauge $\nabla\cdot {\vec a}=0$, satisfying 
$b=\nabla\times {\vec a}=2\pi{\tilde\phi}\rho$.
In deriving ${\vec v}_{\tilde\phi}$, we have used 
$\nabla({\rm Re}\log z)=\nabla
({\rm Im}\log z)\times \hat{n}$.

The physical justification for including the real 
part of $\tilde{\phi}\log (z-z')$ in the CF transformation 
(\ref{J}) lies in the fact that the resulting 
mean-field states give rise to numerically well-tested
wave functions. To see this, we note that at the 
mean field level, one ignores the fluctuations 
of the Chern-Simons field $\vec{a}$ and, for fillings 
at or close to $1/\tilde{\phi}$, take $\vec{a}$ to 
be a classical field determined by eq. (\ref{CS}) 
with a uniform density $n_{e}$, as in CSF theory: 
\begin{eqnarray}
\bar{\rho}(\vec{x})=n_{e}, \;\;
\bar{\vec a}(\vec{x})=(B_{\tilde{\phi}}/2)\hat{n}\times 
{\vec x}={\vec A_{\tilde{\phi}}}(\vec{x}).
\label{MFsol}
\end{eqnarray}
Here one has $\hat{n}\times\bar{\vec a}=
{\vec x}/2l^2_{\tilde{\phi}}$, which results  
in $\bar J+\bar J^\dagger=0$. Substituting 
eq. (\ref{MFsol}) into eq. (\ref{CFHam}), we get 
the mean-field Hamiltonian describing free CF's 
in an effective field $\Delta B$:
\begin{equation}
H_{MF}=-\frac{1}{2m_b}\int d^2x\; \Pi(\vec{x})
(\nabla+i \Delta {\vec A})^2 \Phi(\vec{x}).
\label{MFHam}
\end{equation}
Once we get the CF wave function 
for a mean-field state, $\chi (\vec{x}_1,...,\vec{x}_N)
\equiv \langle 0|\Phi(\vec{x}_1) \ldots{}\Phi(\vec{x}_N)
|MF\rangle$, 
the corresponding electron wave function can be
easily read off as
\begin{eqnarray}
&&\psi_{MF}(\vec{x}_1,...,\vec{x}_N)\equiv
\langle 0|\psi(\vec{x}_1)\ldots{}
\psi(\vec{x}_N)|MF\rangle\nonumber\\
&&\;\;\;=\langle 0|e^{J(\vec{x}_1)}\Phi(\vec{x}_1)\ldots{}
e^{J(\vec{x}_N)}\Phi(\vec{x}_N)|MF\rangle\nonumber\\
&&\;\;\;= \prod_{i<j}(z_i-z_j)^{\tilde\phi}
\exp[-\frac{1}{4l^2_{\tilde\phi}} \sum_i|z_i|^2] 
\; \chi (\vec{x}_1,...,\vec{x}_N)\, ,
\label{MFwf}
\end{eqnarray}
recovering Jain's rule for trial wave 
functions\cite{Jain}. Here we used 
the identity (\ref{2.12}) to move all 
$e^{J(\vec{x}_i)}$ to the left, which act on 
the vacuum $\langle 0|$ yielding 
the Gaussian factor. The mean-field CSF 
theory missed the factor $\prod_{i<j} 
|z_i-z_j|^{\tilde\phi}$, whose presence 
in eq. (\ref{MFwf}) is due to the inclusion of
the real part of $\tilde{\phi} \log (z-z')$ 
in our CF transformation (\ref{J}).

At exactly $\nu=1/\tilde{\phi}$, CF's are in zero 
effective magnetic field $\Delta B=0$. The mean-field 
Hamiltonian (\ref{MFHam}) requires the 
ground state be a filled Fermi sea of CF's, with 
Fermi wave vector $k_F=(2\pi\tilde\phi n_e)^{1/2}
=1/l_{\tilde\phi}$. In terms of the CF field in 
momentum space, 
the Fermi-sea state ket is 
\begin{equation}
|G_0\rangle = \prod _{k<k_F}
\Pi(\vec{k}) \; |0\rangle.
\label{FSKet}
\end{equation}
Thus, with $\chi_{0}=
\langle 0|\Phi(\vec{x}_1)
\ldots{}\Phi(\vec{x}_N)|G_{0}\rangle
= \det (e^{i\vec{k}_i\cdot\vec{x}_j})$,
eq. (\ref{MFwf}) reproduces the (unprojected) 
Rezayi-Read trial wave function\cite{RR}. 
According to eq. (\ref{MFHam}), the mean-field 
quantum Hall state at $\nu=p/(\tilde{\phi}p+1)$ 
is the state with CF's completely filling 
$p$ Landau levels in $\Delta B$. 
For $p=1$,
\begin{equation}
\chi_1(\vec{x}_1,\cdot,\vec{x}_N)=\prod_{i<j}(z_i-z_j)
\exp[-\frac{1}{4l^2_{\Delta  B}}\sum_i |z_i|^2]; 
\label{wfchi}
\end{equation}
the mean-field electron wave function (\ref{MFwf})
just gives the Laughlin trial wave function \cite{Lau}
(since $l^{-2}_{\tilde{\phi}}+l^{-2}_{\Delta B}
= l^{-2}_{B}$)
\begin{equation}
\prod_{i<j}(z_i-z_j)^{\tilde{\phi}+1} \,
\exp[-\frac{1}{4l^2_B}\sum_i|z_i|^2].
\label{3.15}
\end{equation}

To systematically improve the mean-field theory,
we do perturbation theory to include the effects
of fluctuations of the Chern-Simons field 
$\vec{a}$. For definiteness, we restrict to the 
filling $\nu=1/\tilde{\phi}$, for which 
$\Delta A=0$. Instead of eq. (\ref{CFHam}), 
we consider the following {\it Hermitian} Hamiltonian: 
\begin{eqnarray}
H_{0}&+&H_{1}= \frac{-1}{2m_b}\int d^2x
\,\Pi(\vec{x})[ \nabla - i(\delta{\vec a}
+i\hat{n}\times \delta{\vec a})]^{2}\,\Phi(x)\nonumber\\
&&+\frac{1}{8\pi \tilde{\phi}^{2}}
\int d^2x d^2x'\; \delta b (\vec{x})\, 
V(\vec{x}-\vec{x}')\, \delta b (\vec{x}').
\label{IntHam}
\end{eqnarray}
with $H_0=H_{MF}$ and $\delta \vec{a} = \vec{a} 
- \bar{\vec{a}}$. We go from the Schr\"{o}dinger 
to the interaction picture by a similar 
but {\it non-unitary} transformation
($H_0$ is not Hermitian): 
\begin{equation}
|\psi(t)\rangle_I
=e^{iH_0t}|\psi(t)\rangle_S, \;
\hat{O}_I(t)=e^{iH_0 t}\,
\hat{O}_S\, e^{-iH_0t}.
\label{IntPic}
\end{equation} 
The CF operators $\Pi_{I}(\vec{x},t)$ 
and $\Phi_{I}(\vec{x},t)$ satisfy the 
usual canonical equal-time anti-commutation 
relations. The evolution operator, 
$U(t,t')\equiv e^{iH_0 t}e^{-iH(t-t')}
e^{-i H_0 t'}$, is no longer
unitary, but we still have 
$U(t,t)=1$, and $U(t,t'')U(t'',t')=U(t,t')$, 
$U^{-1}(t,t')=U(t',t)$. Also the 
Schr\"{o}dinger equation and the Dyson formula 
for $U(t,t')$ are formally the same as before:
\begin{equation}
U(t',t)=T\,\exp\biggl(-i\int_t^{t'} 
d\tau H_1(\tau) \biggr),
\label{Dyson}
\end{equation}
where $H_1(t)\equiv e^{iH_0 t}H_1e^{-iH_0t}$, 
and $T$ does time-ordering.

To proceed, first we need an appropriate basis 
to evaluate the matrix elements of eq. (\ref{Dyson}). 
The base kets and bras for CF's with definite 
momenta \cite{comm1} are given by  
\begin{equation}
|\{\vec{k}_i\}\rangle\equiv
\prod_i \Pi (\vec{k}_i)\, |0\rangle, \;\;\;
\tilde{\langle \{\vec{k}_i\}|} \equiv
\langle 0|\, \prod_i \Phi (\vec{k}_i).
\end{equation}
They are eigenvectors of $H_{0}$ and have orthonormal 
overlaps. (Note that the bras $\langle \{\vec{k}_i\}|$
do not have orthonormal overlaps with the kets 
$|\{\vec{k}_j\}\rangle$!) 
Thus, corresponding to the Fermi-sea ket 
(\ref{FSKet}), the bra describing the CF Fermi 
sea, that has a unit overlap with it, is
given by
\begin{equation}
\langle \tilde{G}_0| =\langle 0|
\prod _{k<k_F}\Phi (\vec{k}),\;\;
\langle \tilde{G}_0|G_0\rangle=1 .
\label{3.19}
\end{equation}
Both the ket $|G_0\rangle$ and the bra $\langle 
\tilde{G}_0|$ are eigenvectors of $H_0$ with the 
same energy $\epsilon_0=\sum_{k<k_F}k^2/2m_b$,
while $\langle G_0|$ is not. 
As a rule, corresponding to usual 
expectation values in the unperturbed ground
state, we always consider the matrix elements 
between $\langle \tilde{G}_0|$ and $|G_0\rangle$. 

The free CF propogator is defined as
\begin{equation}
G_0(\vec{x},t;\vec{x}',t')=-i\tilde{\langle G_0|}
T(\Phi_{I}(\vec{x},t) \Pi_{I}(\vec{x}',t'))
|G_0\rangle, 
\label{4.11}
\end{equation}
whose Fourier transform is the same 
as a free electron
\begin{eqnarray}
G_0({\vec k},\omega)=
\frac{\theta(k-k_F)}{\omega-\epsilon_{\vec{k}}+i0^+}+
\frac{\theta(k_F-k)}{\omega-\epsilon_{\vec{k}}-i0^+},
\label{4.13}
\end{eqnarray}
where $\theta(k)$ is the step function and 
$\epsilon_{\vec{k}}=k^2/2m_b$. Introducing $a_{0}$
to implement the field-density constraint,
a Lagrangian approach for the Chern-Simons propogator 
\begin{equation}
D^0_{\mu\nu}(\vec{x},t;0,0)
=-i\langle \tilde{G}_0|T(\delta a_\mu(x,t)
\delta a_\nu(0,0))|G_0\rangle\, ;
\end{equation}
leads to the one in usual CSF theory 
\begin{eqnarray}
D^0_{\mu\nu}({\vec q, \omega})=
\delta (\omega) U_{\mu\nu}(\vec{q}), 
\begin{array}{ccc}
U=\left(\begin{array}{cc}{v({\vec q})}&{\displaystyle
          \frac{2\pi i\tilde\phi}{q}}\\
          {\displaystyle-\frac{2\pi i\tilde\phi}{q}}&{0}
         \end{array}\right).
\end{array}
\label{CSPro}
\end{eqnarray}
Here we have adopted the $2\times 2$ matrix 
formalism\cite{HLR} with $\mu,\nu = 0,1$; $0$ 
stands for the time component, $1$ the space
component transverse to $\vec{q}$. Eq. (\ref{IntHam})
implies a {\it complex} CF-Chern-Simons coupling 
$\rho \delta a_{0}+ (\vec{j}-i\hat{n}\times 
\vec{j})\cdot \delta \vec{a}$, with 
$\vec{j} = (-i/2m_{b})[(\nabla \Pi) \Phi
-\Pi \nabla \Phi]$,  
resulting in the CF-CF-$\delta a$ vertex 
(Fig. 1a)
\begin{equation}
g_\mu (\vec{k}+\vec{q},\vec{k})=\biggl(1,
\frac{(2\vec{k}+\vec{q})\times{\hat q}
+i(2\vec{k}+\vec{q})\cdot{\hat q}
}{2m_b}\biggr),
\label{4.16}
\end{equation}  
where $\hat{q}=\vec{q}/q$ is a unit vector.
Note that the second term in the $\mu=1$ 
component is absent in the CSF theory. 
Moreover, in view of $(\delta {\vec a} 
+i\hat{n}\times\delta {\vec a})^2=0$, 
there is no CF-CF-$\delta a$-$\delta a$ vertex
(Fig. 1b) in our Feynman rules. This makes the 
structure of Feynman diagrams more like a 
theory with two-body potential than  
usual gauge theory. 

Using the Wick theorem and the above Feynman rules, 
one can calculate as usual the matrix elements of 
the evolution operator (\ref{Dyson}) between free 
CF states. To make connection to physics, 
we have managed to prove a generalized Gell-Mann-Low 
theorem, which relates these matrix elements 
to correlation functions: 
\begin{eqnarray}
&& \langle G|T O_{1,H} (\vec{x}_1,t_1) \cdots
O_{n,H} (\vec{x}_n,t_n) |G\rangle \nonumber\\ 
= && \frac{\langle \tilde{G_0}|T \prod_{i} O_{i,I} 
(\vec{x}_i,t_i) U (\infty, -\infty) |G_0 \rangle}
{\langle \tilde{G_0}| U (\infty, -\infty) 
|G_0 \rangle}\, ,
\label{GGL}
\end{eqnarray}
where $|G\rangle$ is the true ground state, 
$O_{i,H} (\vec{x}_i,t_i)$ are local operators in terms of
$\Phi_H$ and $\Pi_H$ in the Heisenberg picture. 
The result, central to our paper, implies a consistent 
perturbation theory with usual diagrammatic techniques, 
despite nonunitarity of our CF transformation 
(\ref{PhiPi}). The proof follows the same steps 
as in usual many-body perturbation theory \cite{Noz}, 
but we have to be careful about problems 
due to non-Hermiticity of $H_0$ and $H_1$.

The basis of the theorem (\ref{GGL})
lies in the following lemma: The state 
obtained from the free CF Fermi-sea 
$|G_0\rangle$ by adiabatically 
switching on $H_{1}(t)$, 
\begin{equation}
|G\rangle\equiv C \lim_{\eta\to 0+} 
\frac{U_{\eta}(0,-\infty)\, |G_0\rangle}
{\langle\tilde{G_{0}}|U_{\eta}(0,-\infty)|G_0\rangle}
\label{GPhys}
\end{equation}
is an eigenstate of $H_{0}+H_{1}$ which, 
by the adiabatic hypothesis, 
is assumed to be the true ground state of 
the system. Here $C$ is a normalization constant
and $U_{\eta}(t,t')$ the 
operator (\ref{Dyson}) with $H_{1}(\tau)
\to e^{-\eta|\tau|}H_{1}(\tau)$.

To make sense, the limit in the right side 
of eq. (\ref{GPhys}) has to exist. We first 
note that, by purely combinatoric considerations
as usual, $U_{\eta}=U_{\eta L}\exp(U_{\eta 0c})$, 
where $U_{\eta L}$ is the linked part of $U_{\eta}$ , 
while $U_{\eta 0 c}$ the sum of the contributions 
to $\langle \tilde{G}_0|U_{\eta}|G_0\rangle$ 
from all unlinked connected diagrams.
It can be shown that as $\eta \to 0+$,
$U_{\eta L}(0,-\infty)$ is regular, while 
$U_{\eta 0c}(0, -\infty)$ diverges as $1/\eta$
due to the integration of $e^{\eta t_i}$ 
in eq. (\ref{Dyson}): 
\begin{eqnarray}
U_{\eta 0c}(0,-\infty)=iA/\eta+\ln C,
\label{4.20}
\end{eqnarray}
where $A$ is $\eta$-independent.
The potential problem lies in the possibility
that $A$ may have a non-zero imaginary part;
then the state $U_\eta(0,-\infty)|G_0\rangle$ 
would have either zero or divergent norm as 
$\eta\to 0+$, making the limit in eq. (\ref{GPhys}) 
nonsense. Usually $U_{\eta}$ is unitary, so $A$ is real. 
However, this argument does not apply in our case, 
since our $H_{1}$ contains a complex Chern-Simons
coupling for CF's. We have managed to check 
explicitly up to three loops, and to give an 
argument for any number of loops, that diagram by 
diagram the contribution to $A$ is real. Thus 
eq. (\ref{4.20}) gives rise to a harmless divergent 
phase factor to the numerator in eq. (\ref{GPhys}), 
that is cancelled by the denominator, and ensures 
the perturbative stability of the mean-field 
Fermi-sea ground state $|G_0\rangle$. 
Moreover, we have been careful to make 
sure that the usual proof for the state
(\ref{GPhys}) to be an eigenstate of
$H_{0}+H_{1}$ goes through in our case:
Only the commutation relations are needed 
here; whether $H_0$ and $H_1$ are
Hermitian is irrelevant.

Similarly, by adiabatically switching off 
$H_1(t)$, we get
\begin{equation}
\langle G| \equiv C \lim_{\eta\to 0+} 
\frac{\langle\tilde{G_0}|\, U_{\eta}(\infty,0)}
{\langle\tilde{G_{0}}|U_{\eta}(\infty,0)|G_0\rangle}\, ,
\label{GPhys2}
\end{equation}
and $U_\eta (\infty, -\infty)\,
|G_0\rangle = \exp (i 2A/\eta) |G_0\rangle $, again
ensuring the perturbative stability of 
$|G_0\rangle$. With these results, one easily
arrives at the generalized 
Gell-Mann-Low theorem. 

In addition to the formal development, we have 
calculated the density-current response functions, 
$K_{\mu\nu} (\vec{q},\omega)$, in the RPA\@ in 
appropriate limits, to explicitly check that 
our formalism indeed gives reasonable results.
The relevant Feynman diagrams are given in Fig. 2. 
We note that our Feynman rules, compared to  
usual CSF theory, have modified the Chern-Simons 
but not the electromagnetic couplings. Besides, 
the RPA equation for our $K_{\mu\nu}$ is different 
from usual CSF theory. Explicit calculations 
shows that the effects of these two differences 
cancel for $K_{\mu\nu}$ both in the static 
($\omega=0$) and high-frequency ($\omega 
\gg qk_F/m_b$) limits for small wavevector 
$q\ll k_F$. We may summarize the differences 
between our theory and usual CSF theory by 
rewriting the RPA equation for our $K_{\mu\nu}$ 
in the following form:
\begin{equation}
K=K^0-K^0[K^0+\Delta K -U^{-1}]^{-1}K^0,
\label{Diff}
\end{equation}
where $K^0$ is the response function of 
the noninteracting CF system (see Fig. 2), 
governed by $H_0$; the matrix $\Delta K$,
absent in usual CSF theory, is a diagonal 
$2\times 2$ matrix 
\begin{equation}
\Delta K = {\rm diag}\biggl(0,\hat{K}^0_{11}
-K^0_{11}-(\hat{K}^0_{01})^2/\hat{K}^0_{00}\biggr).
\end{equation}
Here $\hat{K}^0$ is the response function of 
non-interacting CF's with two Chern-Simons vertices
(Fig. 2); in deriving eq. (\ref{Diff}) we exploited 
relations between the explicit expressions of
the one-loop blocks in Fig. 2. Straightforward
calculation shows that both in the static
($\omega=0$) and high-frequency ($\omega \gg 
qk_F/m_b$) limits for small wavevector $q\ll k_F$, 
$\Delta K$ just happens to vanish: 
\begin{equation}
\hat{K}^0_{11}- 
(\hat{K}^0_{01})^2/\hat{K}^0_{00}= K^0_{11}.
\label{NoDiff}
\end{equation}
Therefore, for long-wavelength fluctuations 
either in the static or high-frequency limit,
there is no physical difference between our 
theory and CSF theory for the linear-response 
functions. This provides a consistency 
check of our perturbation theory: 
On one hand, the size of the vortex core in 
a CF is of the order of the magnetic length, so 
a probe with wavelength much bigger than that 
should not be able to see the core; on the 
other hand, the high-frequency behavior of the 
response function should be constrained by Kohn's 
theorem\cite{Kohn} (with mass $m_b$), 
which is indeed satisfied by our RPA results. 
Thus, the experimental predictions for probes with
long wavelength discussed 
by Halperin, Lee and Read\cite{HLR} 
remain unchanged. Furthermore, when higher-order 
contributions are included, the ``Fermi-liquid corrections'' 
 \cite{SSH} are needed and can be done as well. 
The details will be given elsewhere\cite{WuYu}.

Nevertheless, there is no reason to believe 
that eq. (\ref{NoDiff}) could be generally 
true. To uncover the new physics due to
the finite vortex core in a CF, which is 
expected to show up for shorter wavelengths 
or larger wave vectors, one needs to 
evaluate the one-loop integrals with 
parameters in certain intermediate range.
Work is in progress. 

To conclude, several comments are in order. First, 
our perturbation formalism can be easily 
generalized to study the incompressible fractional 
quantum Hall fluids either from the CF scenario
or from the composite boson scenario; in 
the latter the integer $\tilde{\phi}$ in eq. 
(\ref{J}) is taken to be odd, resulting a 
canonical boson pair $\Phi$ and $\Pi$ \cite{Raj}.
Here we have considered an infinite, homogeneous
system without boundary. It would be 
interesting to generalize the present 
formulation to a finite compact geometry, 
say a sphere or a torus, which may be helpful
for clarifying some consequences of the 
non-orthogonality of CF states, such as fractional 
and mutual exclusion statistics between quasiholes 
and quasielectrons \cite{Hald2,Wu}.
Also it is worth to study the effects due to a finite 
vortex core for quasiparticles on the edge of a 
finite system with boundary.
%from the point of view of the bulk-edge interplay.

Y.S.W. would like to thank F.D.M. Haldane
for discussions. The work was supported in 
part by grant NSF PHY-9309458.


\begin{references}
\bibitem{comm1}
In this paper we consider an infinite, homogeneous
bulk system without disorder.
\end{references}
\end{document}